\documentclass{article}
\usepackage{arxiv}

\usepackage[utf8]{inputenc}
\usepackage[T1]{fontenc}
\usepackage{xcolor}
\usepackage{amsmath}
\usepackage{amssymb}
\usepackage{enumitem}
\usepackage{makecell}
\usepackage{booktabs}
\usepackage{theorem}
\usepackage{hyperref}
\usepackage{graphicx}
\usepackage{listings}
\usepackage{multirow}
\usepackage{cleveref}
\lstdefinestyle{GherkinStyle}{
    language=Ruby,
    basicstyle=\ttfamily\footnotesize,
    keywordstyle=\color{blue},
    stringstyle=\color{red!70!black},
    commentstyle=\color{gray},
    morekeywords={Given, When, Then, do, end},
    frame=single,
    rulecolor=\color{lightgray},
    breaklines=true,
    breakatwhitespace=true,
    morekeywords={Scenario, Outline, And, Given, When, Then, Examples, Formal}
}

\definecolor{sv_keyword}{rgb}{0.13, 0.13, 1}
\definecolor{sv_type}{rgb}{0, 0.5, 0}
\definecolor{sv_string}{rgb}{0.64, 0.08, 0.08}
\definecolor{sv_comment}{rgb}{0.5, 0.5, 0.5}

\lstdefinestyle{VerilogStyle}{
    language=Verilog,
    basicstyle=\ttfamily\footnotesize,
    keywordstyle=\color{sv_keyword}\bfseries,
    commentstyle=\color{sv_comment}\itshape,
    stringstyle=\color{sv_string},
    showstringspaces=false,
    numbers=left,
    numberstyle=\tiny\color{gray},
    numbersep=8pt,
    frame=lines,
    rulecolor=\color{lightgray},
    breaklines=true,
    tabsize=2,
    morekeywords={
        logic, bit, byte, int, longint, shortint, string,
        struct, enum, union, type, typedef,
        class, extends, new, virtual, local, protected,
        package, import, export, interface, modport,
        program, clocking, default, input, output, inout,
        ref, const, static, automatic,
        foreach, return, break, continue,
        assert, property, sequence, cover,
        rand, constraint, solve, before,
        mailbox, semaphore, event, wait_order
    }
}

\lstdefinestyle{SVAStyle}{
    style=VerilogStyle,
    keywordstyle=[2]\color{purple}\bfseries,
    keywordstyle=[3]\color{teal}\bfseries,
    morekeywords=[2]{
        assert, property, sequence, cover, assume, restrict,
        expect, disable, iff, first_match, intersect,
        throughout, within, within_order, and, or, not
    },
    morekeywords=[3]{
        posedge, negedge, edge, @, 
        \#\#0, \#\#1, \#\#2, |->, |=>, [*], [=], [->]
    },
    frame=single
}

\definecolor{prompt_bg}{rgb}{0.97, 0.97, 1.0}
\definecolor{prompt_var}{rgb}{0.7, 0.1, 0.1}
\definecolor{prompt_instr}{rgb}{0.2, 0.2, 0.2}

\lstdefinestyle{PromptStyle}{
    basicstyle=\ttfamily\footnotesize\color{prompt_instr},
    backgroundcolor=\color{prompt_bg},
    breaklines=true,
    breakatwhitespace=true
    frame=leftline,
    framerule=2pt,
    rulecolor=\color{blue!30},
    xleftmargin=0pt,
    framesep=0pt,
    morekeywords={System, User, Assistant, Instruction, Context, Output},
    keywordstyle=\bfseries\color{blue!60!black}
}

\usepackage[nolist]{acronym}

\begin{acronym}
    \acro{eda}[EDA]{Electronic Design Automation}
    \acro{rtl}[RTL]{Register Transfer Level}
    \acro{llm}[LLM]{Large Language Model}
    \acro{fpv}[FPV]{Formal Property Verification}
    \acro{bdd}[BDD]{Behavior Driven Development}
    \acro{cnl}[CNL]{Controlled Natural Language}
    \acro{fvgherkinscenario}[FV Gherkin Scenario]{Formal Verification Gherkin Scenario}
    \acro{fvgherkinspecification}[FV Gherkin Specification]{Formal Verification Gherkin Specification}
    \acro{genai}[GenAI]{Generative Artificial Intelligence}
    \acro{sva}[SVA]{SystemVerilog Assertion}
    \acro{nlp}[NLP]{Natural Language Processing}
    \acro{hdl}[HDL]{Hardware Description Language}
    \acro{alu}[ALU]{Arithmetic Logic Unit}
    \acro{cpu}[CPU]{Central Processing Unit}
    \acro{api}[API]{Application Programming Interface}
    \acro{fsm}[FSM]{Finite State Machine}
    \acro{lifo}[LIFO]{Last In First Out}
    \acro{ip}[IP]{Intellectual Property}
    \acro{coi}[COI]{Cone of Influence}
    \acro{cot}[COT]{Chain of Thought}
    \acro{rag}[RAG]{Retreival Augmented Generation}
\end{acronym}

\theoremstyle{definition}
\newtheorem{definition}{Definition}
\newtheorem{example}{Example}
\newcommand*{\step}[1]{\textcircled{#1}}

\title{LLM-enabled Behavior Driven Development Workflow for Formally Verified Hardware Designs}
\author{Luca Müller\\ DFKI GmbH \\ Bremen, Germany \And Qian Liu\\ University of Bremen \\ Bremen, Germany \And Rolf Drechsler\\ University of Bremen/DFKI \\ Bremen, Germany}
\date{}

\begin{document}
	
\maketitle

\acresetall

\begin{abstract}
	Recently, the use of \acp{llm} for different tasks in the \ac{eda} life-cycle has been studied extensively, but an integrated view is lacking.
	Specifications are the foundation of this life-cycle, but they suffer from ambiguity when written in natural language, which especially affects the quality of \ac{llm} output.
	Formal specifications mitigate these ambiguities, but they come with their own challenges.
	On the other hand, \ac{cnl} specifications can serve as a middle-ground, reducing ambiguity while retaining interpretability.
	
	In this work, we propose an integrated view on the use of \acp{llm} for \ac{eda} and establish an \ac{llm}-enabled behavior driven hardware development workflow.
	We introduce and define \acp{fvgherkinscenario}, unlocking \ac{cnl} specifications as the foundation for formally verified hardware designs via \ac{fpv}.
	Experimental evaluation shows that our workflow is able to outperform other established \ac{llm}-based methods by 2.48x in functional correctness of generated \ac{rtl} designs and by 2.54x in formal coverage of generated assertions for \ac{fpv}.
\end{abstract}


\section{Introduction} \label{sec:intro}

\acp{llm} have been a major focus of research since their advent some years ago and are being explored throughout various \ac{eda} tasks~\cite{llm_for_eda}, including \ac{rtl} design and assertion generation for \ac{fpv}~\cite{assertionforge}.
While some works review multiple parts of the \ac{eda} flow, we identify the lack of an integrated view on the use of \acp{llm} for these tasks~\cite{llm_for_eda,llm_for_verification_test_design}.
Ideally, we envision an integrated workflow spanning the entire \ac{eda} life-cycle, where agentic \acp{llm} autonomously carry out streamlined processes under human supervision to produce hardware designs that are formally verified to provide correctness guarantees.
Specifications are the foundation of this workflow, serving as the ground-truth for expected hardware functionality and behavior.
Today, specifications are typically written in natural language, which inherently carries ambiguity that can lead to unexpected results~\cite{ambiguity_nlp}, especially for \acp{llm}, whose output heavily depends on the input format~\cite{prompt_format}.
Previous works aim to mitigate this ambiguity by utilizing techniques like \ac{cot}~\cite{cot} prompting or \ac{rag}~\cite{rag} to guide \ac{nlp} of \acp{llm}, but work that targets the refinement of the specification itself is sparse~\cite{specllm}.
In this regard, formal specifications are one potential solution, but they can be difficult for humans to understand~\cite{formal_spec_barriers}, hindering proper audition of \ac{llm}-generated artifacts.
\ac{cnl}~\cite{cnl} specifications pose a promising alternative, serving as a middle-ground and reducing ambiguity with their structure while retaining easy human interpretability through their roots in natural language.
Initial attempts at \ac{cnl} specifications for hardware design have been explored, e.g., PROSER for temporal model checking~\cite{prosper} and Gherkin embedded in \ac{bdd} for hardware design~\cite{bdd_hardware}.
However, practical adoption of \ac{cnl} specifications remains scarce, which we attribute to limitations with respect to expressiveness beyond temporal model checking and generalizability to \ac{fpv} respectively.

In this work, we propose an integrated view on the use of \acp{llm} for \ac{eda} with the establishment of an \ac{llm}-enabled behavior driven hardware development workflow.
Our workflow directly incorporates \ac{fpv} evaluation tools, opening an end-to-end path from design requirements to formally verified hardware designs.
We introduce and define \acp{fvgherkinscenario}, unlocking the potential of \ac{cnl} specifications as the foundation of this integrated flow.
Experimental evaluation confirms the effectiveness of our proposed workflow, demonstrating that it outperforms other established \ac{llm}-based methods on contemporary benchmarks in both \ac{rtl} design and assertion generation tasks.
Specifically, we make the following contributions:

\begin{itemize}
	\item Present an \ac{llm}-enabled behavior driven hardware development workflow, integrating specification, \ac{rtl} design, and formal verification stages.
	\item Introduce \acp{fvgherkinscenario} as a specification baseline for \ac{rtl} design and formal verification.
	\item Demonstrate how the integration of \acp{llm} into our workflow can increase productivity and improve verification quality.
\end{itemize}

The remainder of this work is structured as follows.
Section~\ref{sec:prelim} introduces preliminary information to keep this work self-contained.
Section~\ref{sec:workflow} presents our proposed workflow and introduces \acp{fvgherkinscenario}.
Section~\ref{sec:exp} experimentally evaluates our workflow, demonstrating its effectiveness.
Section~\ref{sec:conclusion} concludes this work and provides an outlook for future research directions.

\section{Preliminaries} \label{sec:prelim}

\subsection{Behavior Driven Development}

\ac{bdd}~\cite{bdd} was originally introduced as an extension to test driven development~\cite{tdd} in the software world. 
The core idea is to start development from a specification of system behavior in a \ac{cnl}, commonly referred to as Gherkin~\cite{gherkin}. 
Multiple scenarios describe how the system behaves under a given set of conditions and are written in a \emph{Given-When-Then} style, structuring pre-conditions, trigger conditions, and post-conditions. 
Logical operators increase the expressiveness of scenarios.
In addition, a general feature overview and background information may be given. 
The adoption of \ac{bdd} for hardware design was initially proposed by Diepenbeck et al.~\cite{bdd_hardware}.
\begin{lstlisting}[style=GherkinStyle, caption=Specification of ADD operation in Gherkin syntax, label=lst:gherkin_scenario, float=ht, belowskip=-1em]
@add @arithmetic
  Scenario Outline: Specify ADD operation
    Given the reset signal rst is low
    And I have operand A = <A>
    And I have operand B = <B>
    And the opcode is set to 0000
    When a rising edge occurs on the clk signal
    Then the result should be <Expected_Result>
    And the zero flag should be <Zero_Flag>
    And the overflow flag should be <Overflow>
    And the negative flag should be <Negative_Flag>

    Examples:
    ...
\end{lstlisting}

\Cref{lst:gherkin_scenario} shows an example of a Gherkin scenario that specifies the addition operation of an \ac{alu}. 
The tags prefixed with $@$ annotate the scenarios according to their purpose in the specification.
The \emph{Scenario Outline} keywords specify that multiple values are tested by this scenario and are followed by a name for the scenario.
After this metadata, the scenario definition follows in \emph{Given-When-Then} style.
The first pre-condition specifies that the reset signal should be low to ensure normal operation.
With the help of the logical operand \emph{And}, operand \emph{A} is specified to have concrete value \emph{<A>} and operand \emph{B} is specified to have concrete value \emph{<B>}.
Additionally, the opcode should be set to constant \emph{0000} for the addition operation.
Next, the trigger condition is specified with keyword \emph{When}, which signifies that the operation should occur on a rising clock edge.
Finally, post-conditions that should hold after the trigger condition are specified with keyword \emph{Then}.
Here, expected values for all outputs are specified by placeholders \emph{<Expected\_Result>}, \emph{<Zero\_Flag>}, \emph{<Overflow>}, and \emph{<Negative\_Flag>}.
After the definition of a scenario outline, an example table defines test cases in the form of values for each of the defined placeholders.
Scenarios in this style are defined until the given hardware has been exhaustively specified.
As evident from \Cref{lst:gherkin_scenario}, this implicit specification by example data imposes limitations regarding the generalizability of Gherkin scenarios to assertions for \ac{fpv}.
Another downside of the current \ac{bdd} approach introduced in ~\cite{bdd_hardware} is the high manual effort required to write scenario definitions for the complete system behavior.
To combat this, first attempts to integrate \acp{llm} for \ac{bdd}-based hardware development were recently explored~\cite{bdd_llm}.
However, formal verification was not targeted, and no end-to-end workflow was proposed here.

\subsection{Large Language Models for Electronic Design Automation}

\begin{table}[t]
\centering
\caption{Comparison to related work}
\label{tab:related}
\resizebox{.7\linewidth}{!}{
\begin{tabular}{l||c|c|c|c|c|c|c|c||c}
\hline
\textbf{Work} & \cite{bdd_llm} & \cite{llm_for_verification_test_design} & \cite{specllm} & \cite{lisa} & \cite{assertllm2} & \cite{llm_for_eda} & \cite{codev} & \cite{mage} & \textbf{Ours} \\ \hline \hline
\textbf{Spec?} & $\checkmark$ & $\times$ & $\checkmark$ & $\checkmark$ & $\times$ & $\times$ & $\times$ & $\times$ & $\checkmark$ \\ \hline
\textbf{RTL?} & $\checkmark$ & $\checkmark$ & $\times$ & $\times$ & $\times$ & $\checkmark$ & $\checkmark$ & $\checkmark$ & $\checkmark$ \\ \hline
\textbf{FPV?} & $\times$ & $\checkmark$ & $\times$ & $\checkmark$ & $\checkmark$ & $\times$ & $\times$ & $\times$ & $\checkmark$ \\ \hline
\textbf{Workflow?} & $\times$ & $\times$ & $\times$ & $\checkmark$ & $\checkmark$ & $\times$ & $\checkmark$ & $\checkmark$ & $\checkmark$ \\ \hline
\end{tabular}}
\end{table}

\acp{llm} are transformer-based \ac{genai} models designed to generate human language and are based on the self-attention mechanism~\cite{attention}.
They are trained on very large amounts of data, which makes them proficient not only in natural language, but also in several programming and \acp{hdl}~\cite{pybench,autochip}.
\acp{llm} take inputs in the form of prompts, and predict output text based on their training.
On the other hand, agentic \acp{llm}, also referred to as agents, can reason, act, and interact, gaining autonomous access to tools~\cite{agentic}.
\acp{llm} can either be run locally or accessed via an \ac{api}.
Recently, both agentic and non-agentic \acp{llm} have been considered for the automation of several \ac{eda} tasks, including specifications generation~\cite{specllm}, generating \ac{rtl} code~\cite{verilogcoder,mage,codev,hdlcore,rtl++}, and generating test cases~\cite{llm4dv}.
Regarding formal verification, the generation of \acp{sva} in particular has been studied~\cite{assertllm,assertllm2,chiraag,assertionforge,lisa,isvlsi}.
Some works consider the use of \acp{llm} in multiple \ac{eda} tasks, but mostly focus on specific case studies rather than introducing a comprehensive workflow~\cite{llm_for_eda,llm_for_verification_test_design}.
To evaluate the effectiveness and quality of using \acp{llm} in \ac{eda}, several benchmarks have been proposed~\cite{assertllm,fveval,assertionbench,assertllm2,chipbench,cvdp,rtllm2,rtllm,verilogeval}.

\Cref{tab:related} compares our proposed workflow to selected related works.
Specifically, we bridge the current research gap with an integrated view through the establishment of an end-to-end workflow from specification to a formally verified implementation.

\section{Workflow} \label{sec:workflow}

\begin{figure*}
	\includegraphics[width=\linewidth]{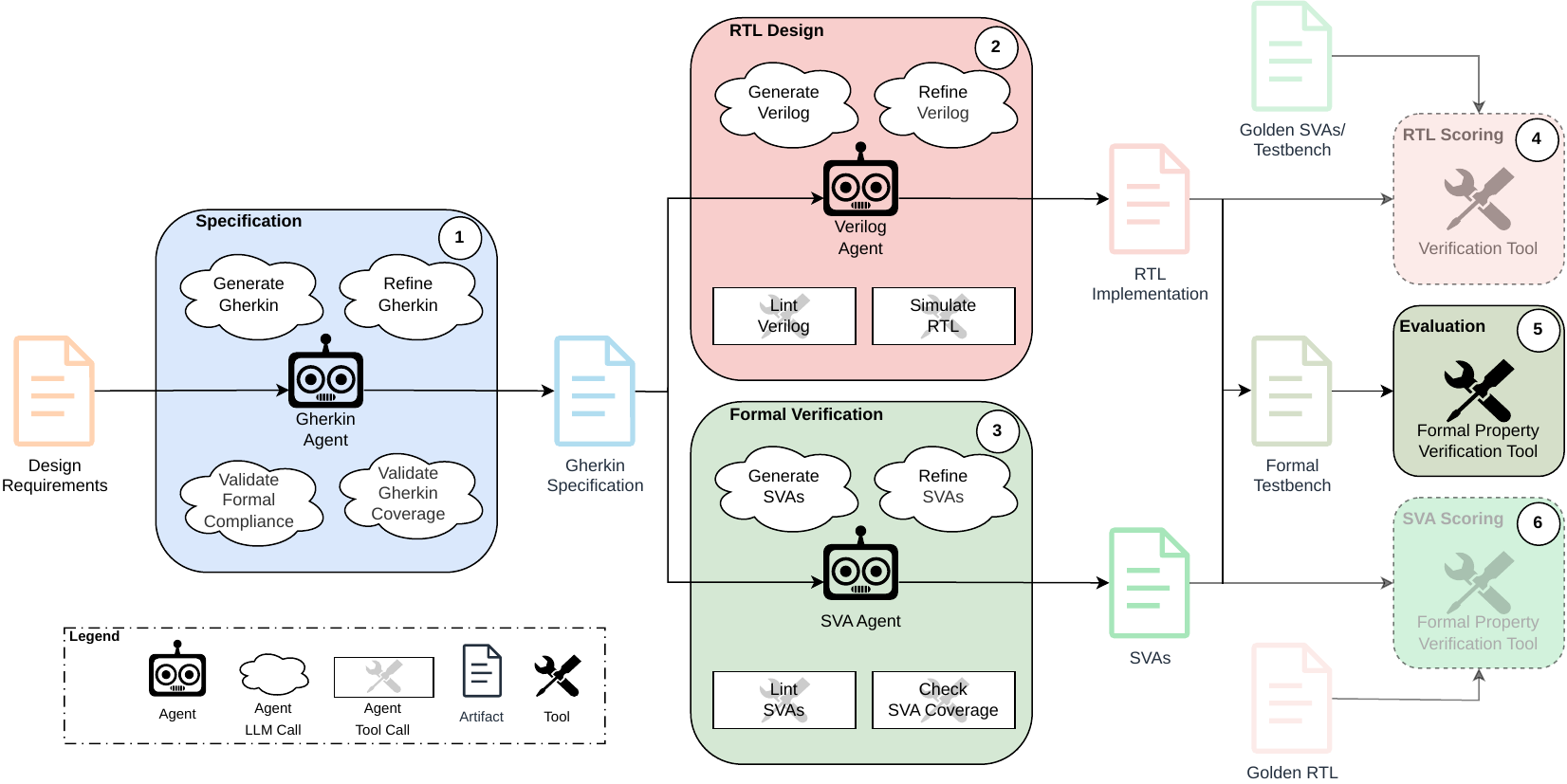}
	\caption{Our proposed \ac{llm}-enabled \ac{bdd} workflow is modeled after the standard hardware development process.}
	\label{fig:workflow}
\end{figure*}

Our proposed \ac{llm}-enabled behavior driven hardware development workflow is presented in \Cref{fig:workflow}.
It features a three agent setup, with specialized agents for specification~\step{1}, \ac{rtl} design\step{2}, and formal verification\step{3}.
Each agent produces an artifact based on its purpose in the workflow, which is conveyed through a system prompt.
Agents can reason, act, and interact with the help of respective \ac{llm} and tool calls at their disposal.
By default, the workflow starts with the input design requirements, which may be provided as plain text, in Markdown, or in PDF format.
From it, the \textit{Gherkin Agent}~\step{1} generates a specification consisting of \acp{fvgherkinscenario}.
Based on this \ac{fvgherkinscenario}, the \textit{Verilog Agent}~\step{2} produces an \ac{rtl} implementation, while the \textit{\ac{sva} Agent}~\step{3} produces a set of \acp{sva}.
Finally, the \ac{rtl} implementation and the \acp{sva} are combined into a formal testbench and evaluated by an \ac{fpv} tool~\step{5}.
Alternatively, the workflow may take the \ac{rtl} path~(\step{1} $\rightarrow$ \step{2} $\rightarrow$ \step{4}) or the \ac{sva} path~(\step{1} $\rightarrow$ \step{3} $\rightarrow$ \step{6}), where only \ac{rtl} implementation or \acp{sva} are generated based on the Gherkin specification, respectively.
In this case, the generated artifacts are scored against a suitable reference provided externally.
In the following, we provide further detail on the three agents and introduce the concept of \acp{fvgherkinscenario}.

\subsection{FV Gherkin Specification}

In the first step of our workflow~\step{1}, design requirements in unstructured natural language are translated to a complete Gherkin specification.
In order to enable formal verification based on the generated specification, we introduce the concept of \acp{fvgherkinscenario}.

\begin{definition} \label{def:fv_gherkin}
    A \textbf{Formal Verification Gherkin Scenario} (\ac{fvgherkinscenario}) must fulfill the following three requirements:
    \begin{enumerate}[label=R\arabic*:, ref=R\arabic*]
        \item\label{req:ex} Any signal value present in the scenario must not be implicitly represented by example data (i.e., no $<>$ placeholders).
        \item\label{req:sym} Any signal value referenced in the post-condition of a scenario must be represented as a symbolic value over signals occurring in the pre-condition, or as a constant.
        \item\label{req:tab} The scenario must not contain an example table for the definition of test cases.
    \end{enumerate}
\end{definition}

Constant values may be required for specific situations, e.g., specifying reset behavior.
In all remaining cases, signal values in the post-condition should be symbolic, i.e., representing the complete value range admissive after the trigger condition.

In our workflow, the Gherkin agent receives the design requirements as input and outputs a specification consisting completely of \acp{fvgherkinscenario}.
To achieve this, it can freely make the following calls in any order:
\begin{itemize}
    \item \textit{Generate Gherkin}: Generate Gherkin scenarios based on design requirements and signal names extracted from them
    \item \textit{Refine Gherkin}: Add, remove, or modify scenarios based on feedback from other \ac{llm} calls
    \item \textit{Validate Formal Compliance}: \ac{llm} judge that checks previously generated scenarios for compliance to FV Gherkin syntax as given by \Cref{def:fv_gherkin}, providing feedback to the agent
    \item \textit{Validate Gherkin Coverage}: \ac{llm} judge that validates the coverage of all design requirements by currently generated scenarios, providing a quality score and feedback on potential coverage holes
\end{itemize}

The agent may choose to converge and commit the current Gherkin specification based on \ac{llm} judge feedback received from calls to \textit{Validate Formal Compliance} and \textit{Validate Gherkin Coverage}.

\begin{lstlisting}[style=GherkinStyle, caption=FV Gherkin specification of ADD operation, label=lst:fv_gherkin, float=h!]
	@add @arithmetic
	FV Scenario: Specify ADD operation
	Given the reset signal rst is low
	And I have operand A
	And I have operand B
	And the opcode is set to 0000
	When a rising edge occurs on the clk signal
	Then the result should be A+B
	And the zero flag should be 1 if A+B=0, else 0
	And the overflow flag should be 1 if the MSB of A and B are equal and the MSB of A+B is different else 0
	And the negative flag should be 1 if the MSB of A+B is 1, else 0
\end{lstlisting}

\begin{example}
    \Cref{lst:fv_gherkin} shows an example of a \ac{fvgherkinscenario} for the specification of the addition operation of an \ac{alu}.
    The \emph{FV Scenario} keywords indicate a \ac{fvgherkinscenario}, in contrast to the \emph{Scenario Outline} to be filled with example data in the specification depicted in \Cref{lst:gherkin_scenario}.
    In the pre-condition, operands \emph{A} and \emph{B} are not specified by concrete values but simply given for use in the post-condition~(\ref{req:ex}).
    The trigger condition stays the same as in \Cref{lst:gherkin_scenario}, whereas the post-condition differs the most.
    In the \ac{fvgherkinscenario}, all four outputs are represented by symbolic values~(\ref{req:sym}).
    For example, the \emph{result} output is specified to be equal to $A+B$ according to the addition operation.
    Instead of dedicating a separate scenario for each combination of binary flags, which can take only values $0$ or $1$, they are also specified symbolically, e.g., the \emph{zero} flag should be $1$ if $A+B=0$ and $0$ otherwise.
    This symbolic definition circumvents the need for an example table~(\ref{req:tab}).
\end{example}

\subsection{RTL Design}

\begin{lstlisting}[style=VerilogStyle, caption=\ac{rtl} design for the given Gherkin specification, label=lst:rtl_design, float=h!]
	module alu (
	input  wire [15:0] a,
	input  wire [15:0] b,
	input  wire [3:0]  opcode,
	input  wire        clk,
	input  wire        rst,
	output reg  [15:0] result,
	output reg         zero,
	output reg         overflow,
	output reg         negative
	);
	...
	always @* begin
	comb_result   = 16'd0;
	comb_overflow = 1'b0;
	...
	case (opcode)
	4'b0000: begin // ADD
	comb_result = a + b;
	comb_overflow = (a[15] == b[15]) && (comb_result[15] != a[15]);
	end
	...
	assign comb_zero     = (comb_result == 16'd0);
	assign comb_negative = comb_result[15];
	
	always @(posedge clk) begin
	if (rst) begin
	...
	end else begin
	result <= comb_result;
	zero <= comb_zero;
	overflow <= comb_overflow;
	negative <= comb_negative;
	end
	end
	endmodule
\end{lstlisting}

The Verilog agent is responsible for \ac{rtl} design~\step{2}, receiving the previously created Gherkin specification as input and producing an \ac{rtl} implementation covering this specification as output.
To do so, it may freely make the following calls:

\begin{itemize}
    \item \textit{Generate Verilog}: Generate Verilog code based on the Gherkin specification, pinned to signal declarations
    \item \textit{Refine Verilog}: Fix Verilog against feedback provided by other calls
    \item \textit{Lint Verilog}: Use an \ac{eda} tool to lint the currently generated Verilog and return feedback on potential syntax errors
    \item \textit{Simulate RTL}: If a reference testbench is available, use an \ac{eda} tool to run it and return feedback on semantic mismatches
\end{itemize}

The Verilog agent converges and commits the \ac{rtl} implementation based on deterministic tool feedback from \textit{Lint Verilog} and \textit{Simulate RTL} calls.

\begin{example} \label{ex:design}
    \Cref{lst:rtl_design} shows an excerpt of the \ac{rtl} design created from the \ac{fvgherkinspecification} for our running \ac{alu} example.
    The design starts with the  module declaration, containing all inputs and outputs specified in the Gherkin specification, taken from the design requirements.
    After some default assignments, operations are defined based on the given \emph{opcode}, modeled after the specific Gherkin scenario that describes its behavior.
    As can be seen, the \emph{result} and \emph{overflow} behavior are directly mapped from the example scenario given in \Cref{lst:fv_gherkin}. 
    The \emph{zero} and \emph{negative} flags can be mapped regardless of the operation, so this behavior is implemented after the case selection on the opcode.
    Finally, the module outputs are assigned synchronously on a rising clock edge, as given in the specification.
\end{example}

As can be observed in \Cref{ex:design}, individual Gherkin scenarios do not necessarily need to be specified as one coherent code block, but may be split across the design.
The important aspect of mapping the Gherkin specification to an \ac{rtl} design is that all scenarios are mapped, and thus the complete behavior of the hardware is implemented.

\subsection{Formal Property Verification}

To enable \ac{fpv} in our workflow, the \ac{sva} agent~\step{3} produces assertions based on the Gherkin specification received as input.
It reaches a set of output \acp{sva} with the help of the following calls:

\begin{itemize}
    \item \textit{Generate \acp{sva}}: Generate \acp{sva} from the \ac{fvgherkinspecification} and signal declarations extracted from previously generated or golden \ac{rtl}
    \item \textit{Refine \acp{sva}}: Refine \acp{sva} against feedback received from other calls
    \item \textit{Lint \acp{sva}}: Elaborate \acp{sva} and a reference \ac{rtl} with an \ac{eda} tool and receive feedback on lint errors
    \item \textit{Check \ac{sva} Coverage}: Evaluate \ac{sva} coverage with an \ac{fpv} tool and receive feedback on coverage holes and failing properties
\end{itemize}

The \ac{sva} agent converges and commits the set of \acp{sva} based on deterministic tool feedback received from calls to \textit{Lint \ac{sva}} and \textit{Check \ac{sva} Coverage}.

\begin{lstlisting}[style=SVAStyle, caption=\ac{sva} for the given Gherkin specification, label=lst:sva, float=h!]
//-------------------
//  Formal Scenario: Perform ADD operation
//-------------------
property p_add;
  @(posedge clk) disable iff (rst)
    (opcode == 4'b0000) |-> ##1
      (result   == (a + b)) &&
      (zero     == ((a + b) == 16'b0)) &&
      (overflow == ((a[15] == b[15]) && (result[15] != a[15]))) &&
      (negative == result[15]);
endproperty
assert property (p_add) else $error("ADD failed");
\end{lstlisting}

\begin{example}
    \Cref{lst:sva} shows an asserted property which verifies the specification of the addition operation given in \Cref{lst:fv_gherkin}.    
    The property directly reflects the structure of the \ac{fvgherkinscenario} with an overlapping implication operator |-> where the antecedent defines the pre-condition and the consequent defines the post-condition.
    For the antecedent, the opcode is set to \emph{0000} according to the pre-condition of the Gherkin scenario to reflect the addition operation.
    The trigger condition of a rising clock edge is achieved by a combination of a clocking event \emph{@(posedge clk)} and a cycle delay \emph{\#\#1}.
    All post-conditions of the \ac{fvgherkinscenario} are directly mapped in the consequent and connected with logical conjunctions \emph{\&\&}.
    Finally, the property is asserted with a fitting error message in case it cannot be satisfied.
\end{example}

After the \ac{sva} agent converges, our end-to-end workflow directly carries out evaluation with the help of an \ac{fpv} tool~\step{5}.
To this end, the generated set of assertions is merged into a formal testbench together with the \ac{rtl} implementation produced by the Verilog agent.
This formal testbench is then executed by the \ac{fpv} tool and filed for further inspection and analysis.

\section{Experimental Evaluation} \label{sec:exp}

We evaluate our workflow on two sets of experiments, each using separate benchmarks and comparing against different baselines.
Firstly, we run the \ac{rtl} path (Requirements $\rightarrow$ \ac{fvgherkinspecification} $\rightarrow$ \ac{rtl} Implementation) to evaluate the quality of \ac{rtl} designs generated based on the \ac{fvgherkinspecification}.
Secondly, in our main experiment, we run the \ac{sva} path (Requirements $\rightarrow$ \ac{fvgherkinspecification} $\rightarrow$ \acp{sva}) to evaluate the quality of generated \acp{sva} and the suitability of \acp{fvgherkinspecification} for \ac{fpv}.
We implement our workflow in Python~\cite{python} and use Cadence Xcelium~\cite{xcelium} for all \ac{rtl} simulation tasks and Cadence JasperGold~\cite{jaspergold} for all \ac{fpv} tasks.
All evaluations are strictly pass@1~\cite{passk} and metrics are retrieved using the aforementioned \ac{eda} tools.
We use the recently released Claude Sonnet 5~\cite{sonnet} as the model for all workflow runs and all baselines where applicable.
For both sets of experiments, we first pick a benchmark that fits our workflow structure, features sufficiently complex designs at an acceptable scale and has baselines available that we can compare against.
Then, we pick suitable baselines that are scorable against the previously selected benchmarks, must be re-scorable to deal with \ac{llm} non-determinism, and are model-adaptable to avoid bias by the choice of \ac{llm} in the best case.
In the following sections, we provide rationale on the benchmarks and baselines we pick to obtain the most realistic evaluation of our workflow.
Chosen benchmarks and baselines are highlighted in boldface.
We note that the given publication dates refer to the publication of peer-reviewed works where applicable, and else, the first mention in an otherwise published work.
All evaluation artifacts are made available.\footnote{\url{https://github.com/lucamu/LLM-BDD-Workflow-Evaluation-Artifacts}}

\subsection{\ac{rtl} Design with our \ac{bdd} Workflow}

\subsubsection{Benchmarks \& Baselines}

\begin{table}[ht]
\centering
\caption{RTL Path Benchmarks}
\label{tab:rtl_bench}
\resizebox{.7\linewidth}{!}{
\begin{tabular}{l||c|c|c|c}
\hline
\textbf{Benchmark} & \textbf{Publication} & \makecell{\textbf{Complexity of} \\ \textbf{Designs}} & \textbf{Scale} & \makecell{\textbf{Baselines} \\ \textbf{Available?}} \\ \hline \hline
RTLLM 2.0~\cite{rtllm2} & 2025-04 & Single-Module & Medium (50) & $\bullet$ \\ \hline
CVDP~\cite{cvdp} & 2025-06 & \makecell{Non-Commercial+ \\ Commercial} & Medium (63)* & $\odot$ \\ \hline
\textbf{VerilogEval v2~\cite{verilogeval}} & 2025-10 & Single-Module & Large (156) & $\bullet$\\ \hline
\textbf{ChipBench}~\cite{chipbench} & 2026-01 & \makecell{Self-Contain+ \\ Hierarchical+IP} & Medium (45) & $\bullet$ \\ \hline
\end{tabular}}
\begin{minipage}{\linewidth}
    \centering \footnotesize *Agentic CID03+CID05 \\
    \footnotesize $\bullet$: Yes \hspace{1em} $\odot$: Partial \hspace{1em} $\circ$: No
\end{minipage}
\end{table}

\Cref{tab:rtl_bench} compares four popular benchmarks targeted at \ac{rtl} design.
We choose \textit{VerilogEval v2} as the largest scale benchmark that many baselines evaluate on, featuring over $150$ designs to get a large sample size.
To cover more complex hierarchical and \ac{ip} designs, we pick \textit{ChipBench}.
We prefer it over \textit{CVDP}, because it was published more recently, reports baseline evaluations and is simpler to integrate and re-score against.
All baselines we pick also provide testbenches that can be scored against a golden reference \ac{rtl} with Xcelium.

\begin{table}[ht]
\centering
\caption{RTL Path Baselines}
\label{tab:rtl_base}
\resizebox{.8\linewidth}{!}{
\begin{tabular}{l||c|c|c|c|c}
\hline
\textbf{Baseline} & \textbf{Publication} & \textbf{Type} & \textbf{Re-Score?} & \makecell{\textbf{Model} \\ \textbf{Adaptable?}} & \makecell{\textbf{VerilogEval}/ \\ \textbf{ChipBench}} \\ \hline \hline
\textbf{VerilogCoder}~\cite{verilogcoder} & 2025-02 & Workflow & $\bullet$\rlap{*} & $\circ$ & $\bullet$/$\circ$ \\ \hline
\textbf{CodeV}~\cite{codev} & 2025-08 & Model & $\bullet$ & N/A & $\bullet$/$\odot$\rlap{**} \\ \hline
HDLCoRe~\cite{hdlcore} & 2025-08 & Workflow & $\circ$ & $\bullet$ & $\circ$/$\circ$ \\ \hline
RTL++~\cite{rtl++} & 2025-08 & Model & $\circ$ & N/A & $\bullet$/$\circ$ \\ \hline
\textbf{MAGE}~\cite{mage} & 2025-09 & Agentic Workflow & $\bullet$ & $\bullet$ & $\bullet$/$\odot$ \\ \hline
\end{tabular}}
\begin{minipage}{\linewidth}
    \centering \footnotesize *Trivial, ships artifacts \hspace{1em} **Custom Adapter \\
    \footnotesize $\bullet$: Yes \hspace{1em} $\odot$: Partial \hspace{1em} $\circ$: No
\end{minipage}
\end{table}

The baselines available for our \ac{rtl} experiment are compared in \Cref{tab:rtl_base}.
\textit{VerilogCoder} ships all generated artifacts for VerilogEval v2, making it trivial to re-score.
\textit{CodeV} is chosen as a fine-tuned model, making integration with additional benchmarks, especially ChipBench, easier.
For \textit{HDLCoRe} and \textit{RTL++}, unfortunately, no implementation is openly available for re-scoring.
As an additional baseline, we compare against \textit{MAGE}, an agentic workflow that was recently released, scores VerilogEval, and is reported on by ChipBench.
We re-run or re-score all baselines with the same Xcelium flow we use for our workflow, using Sonnet-5 where applicable. 

\subsubsection{Experimental Results}

\begin{table}[htbp]
\centering
\caption{RTL Path Results}
\label{tab:rtl_res}
\resizebox{.8\linewidth}{!}{
\begin{tabular}{l|l|l|r|r}
\toprule
Benchmark & Method & Source & \multicolumn{1}{c|}{Syntax} & \multicolumn{1}{c}{Functional} \\
\hline
\multirow{8}{*}{VerilogEval v2}
 & VerilogCoder & Re-Score     & 96.6\% (142/147)* & 90.4\% (141/156) \\ \cline{2-5}
 & \makecell[l]{VerilogCoder \\ (GPT-4 Turbo)}  & Self-Reported~\cite{verilogcoder}  & - & 94.2\% \\ \cline{2-5}
 & CodeV                & Re-Run      & 85.3\% (133/156) & 35.9\% (56/156)  \\ \cline{2-5}
 & CodeV (pass@1)       & Self-Reported~\cite{codev}**  & - & 59.2\% \\ \cline{2-5}
 & MAGE (Sonnet-5)      & Re-Run & 99.2\% (127/128)*** & 76.3\% (119/156) \\ \cline{2-5}
 & MAGE                 & Self-Reported~\cite{mage} & - & 95.7\% \\ \cline{2-5}
 & \textbf{BDD Workflow} & \textbf{Ours} & 98.7\% (154/156) & 89.1\% (139/156) \\
\hline
\multirow{7}{*}{ChipBench}
 & CodeV                         & Re-Run      & 82.2\% (37/45) & 13.3\% (6/45) \\ \cline{2-5}
 & Opus 4.5 (pass@1)             & \makecell[l]{Reported \\ (LLM)~\cite{chipbench}} & - & 30.7\% \\ \cline{2-5}
 & MAGE                          & \makecell[l]{Reported \\ (Workflow)~\cite{chipbench}} & - & 37.4\% \\ \cline{2-5}
 & MAGE (Sonnet-5)               & Re-Run & 95.8\% (23/24)**** & 31.1\% (14/45)\% \\ \cline{2-5}
 & \textbf{BDD Workflow}         & \textbf{Ours}   & 95.6\% (43/45) & 31.1\% (14/45) \\
\bottomrule
\end{tabular}}
\begin{minipage}{\linewidth}
    \centering \footnotesize *9 designs ship no design \hspace{1em} **CodeV-Verilog-QC (best) \\ ***Remaining 28 produced an error \hspace{1em} ****Remaining 21 produced an error
\end{minipage}
\end{table}

\Cref{tab:rtl_res} shows the results of the experimental evaluation of our \ac{rtl} path.
Column \textit{Syntax} reports the number of designs which have valid Verilog syntax and column \textit{Functional} reports the number of designs which pass the testbench provided by the benchmark with 100\% pass rate.
For the VerilogEval v2 benchmark, our proposed workflow is able to produce valid syntax on 154 out of 156~(98.7\%) of designs, being outperformed only by MAGE on Sonnet-5, which, as a caveat, produced errors on $28$ designs.
As for the functional pass rate on the provided testbench, our workflow reaches almost 90\% and performs similar to our highest re-score on VerilogCoder (90.4\%), a difference which may be neglected due to noise in the \ac{llm} generation.
Compared to our CodeV re-run, our proposed workflow improves the functional pass rate by 2.48x.
Looking at the ChipBench benchmark, our workflow once again scores very high on syntax, with over 95\%.
Regarding the functional pass rate, it performs the same as the MAGE re-run on Sonnet-5 and outperforms both CodeV and the best stand-alone \ac{llm} reported in~\cite{chipbench}.
Overall, the results show that our proposed workflow performs better than or on-par with previously released open-source baselines, indicating a general suitability of \acp{fvgherkinspecification} for \ac{rtl} design.

\subsection{\ac{sva} Generation for \ac{fpv} with our \ac{bdd} Workflow}

\subsubsection{Benchmarks \& Baselines}

\begin{table}[ht]
\centering
\caption{SVA Path Benchmarks}
\label{tab:sva_bench}
\resizebox{.9\linewidth}{!}{
\begin{tabular}{l||c|c|c|c|c}
\hline
\textbf{Benchmark} & \textbf{Publication} & \makecell{\textbf{Req}$\rightarrow$\textbf{SVA?}} & \makecell{\textbf{Complexity of} \\ \textbf{Designs}} & \textbf{Scale} & \makecell{\textbf{Baselines} \\ \textbf{Available?}} \\ \hline \hline
\textbf{AssertLLM}~\cite{assertllm} & 2025-03 & $\bullet$ & Industry IP & Small (20) & $\bullet$ \\ \hline
AssertionBench~\cite{assertionbench} & 2025-04 & $\circ$\rlap{*} & Various & Large (100) & $\odot$ \\ \hline
FVEval~\cite{fveval} & 2025-05 & $\odot$\rlap{**} & Unit-Level IP & Medium (79)*** & $\odot$ \\ \hline 
CVDP~\cite{cvdp} & 2025-06 & $\circ$\rlap{****} & \makecell{Non-Commercial+ \\ Commercial} & Medium (30)***** & $\odot$ \\ \hline
\textbf{AssertLLM2}\cite{assertllm2} & 2026-05 & $\bullet$ & Industry IP & Medium (83) & $\bullet$ \\ \hline
\end{tabular}}
\begin{minipage}{\linewidth}
    \centering \footnotesize *RTL$\rightarrow$SVA \hspace{1em} **Assertion Description$\rightarrow$SVA \hspace{1em} ***NL2SVA-Human \\ ****Test Plan$\rightarrow$SVA  \hspace{1em} *****Agentic CID14 \\
    \footnotesize $\bullet$: Yes \hspace{1em} $\odot$: Partial \hspace{1em} $\circ$: No
\end{minipage}
\end{table}

Suitable benchmarks for the \ac{sva} path are more limited, as can be seen in \Cref{tab:sva_bench}.
Available benchmarks greatly differ in the task they evaluate: \textit{AssertionBench} considers \ac{rtl} code for \ac{sva} generation, \textit{FVEval} provides small assertion descriptions that are translated to \ac{sva} one-by-one, and \textit{CVDP} supplies a testplan for \ac{sva} generation.
On the other hand, \textit{AssertLLM} and \textit{AssertLLM2} both benchmark \ac{sva} generation based on design specifications, provide sufficiently complex designs, and have baselines that evaluate on them.
Both of them also provide a golden \ac{rtl} against which generated \acp{sva} can be evaluated using JasperGold.

\begin{table}[ht]
\centering
\caption{SVA Path Baselines}
\label{tab:sva_base}
\resizebox{.9\linewidth}{!}{
\begin{tabular}{l||c|c|c|c|c}
\hline
\textbf{Baseline} & \textbf{Publication} & \textbf{Type} & \textbf{Re-Score?} & \makecell{\textbf{Model} \\ \textbf{Adaptable?}} & \makecell{\textbf{AssertLLM/} \\ \textbf{AssertLLM2}} \\ \hline \hline
ChIRAAG~\cite{chiraag} & 2024-09 & Workflow & $\bullet$ & $\odot$ & $\circ$/$\circ$ \\ \hline
\textbf{AssertionForge}~\cite{assertionforge} & 2025-08 & Workflow & $\bullet$ & $\bullet$ & $\bullet$/$\circ$ \\ \hline
LISA~\cite{lisa} & 2025-08 & Model & $\circ$ & $\circ$ & $\circ$/$\circ$ \\ \hline
\textbf{AssertLLM2 Plain}~\cite{assertllm2} & 2026-05 & Method & $\bullet$ & $\bullet$ & $\circ$/$\bullet$\rlap{*} \\ \hline
\end{tabular}}
\begin{minipage}{\linewidth}
    \centering \footnotesize *Shipped with Benchmark\\
    \footnotesize $\bullet$: Yes \hspace{1em} $\odot$: Partial \hspace{1em} $\circ$: No
\end{minipage}
\end{table}

\Cref{tab:sva_base} compares potential baselines for the \ac{sva} path.
\textit{ChIRAAG} is publicly available but generates \acp{sva} based on on hand-crafted specifications, while \textit{LISA} is not available for re-scoring at all.
For AssertLLM, \textit{AssertionForge} is a very suitable candidate, as it scores directly on it and is also available to re-run.
AssertLLM2 was only released very recently, and thus, no external baselines natively support it yet.
Instead, we make use of the \textit{AssertLLM2 Plain} method as a baseline, which is provided alongside the benchmark.
Both baselines are re-run on the same model and with the same evaluation flow as our proposed workflow using JasperGold.

\subsubsection{Experimental Results}

\begin{table}[htbp]
\centering
\caption{SVA Path Results}
\label{tab:sva_res}
\resizebox{\linewidth}{!}{
\begin{tabular}{l|l|l|r|r|r}
\toprule
Benchmark & Method & Source & Proof & \multicolumn{1}{c|}{COI} & Formal \\
\hline
\multirow{3}{*}{AssertLLM}
 & AssertionForge (Sonnet-5)& Re-Run      & 19.0\% & 63.0\% & 36.7\% \\ \cline{2-6}
 & AssertionForge (GPT-4o)  & Self-Reported~\cite{assertionforge}* & 18.6\% & 96.2\% & - \\ \cline{2-6}
 & \textbf{BDD Workflow}     & \textbf{Ours}                  & 43.2\% & 96.8\% & 44.1\% \\ \cline{1-6}
\hline
\multirow{4}{*}{AssertLLM2} 
 & AssertLLM2 Plain (Sonnet-5)**    & Re-run       & 67.8\% & 99.9\% & 15.5\% \\ \cline{2-6}
 & AssertLLM2 Plain (Sonnet 4.5)    & \makecell[l]{Self-Reported \\ (Average)~\cite{assertllm2}} & 55.3\% & 65.7\% & 22.4\% \\ \cline{2-6}
 & \textbf{BDD Workflow}**           & \textbf{Ours}                          & 43.1\% & 92.9\%  & 39.5\% \\
\bottomrule
\end{tabular}}
\begin{minipage}{\linewidth}
    \centering \footnotesize *Extrapolated from reported five design subset \hspace{1em} **23 design subset
\end{minipage}
\end{table}

The results of the experimental evaluation of our \ac{sva} path are shown in \Cref{tab:sva_res}.
For the AssertLLM2 benchmark, we score a subset of 23 designs, with at least one design per category~\cite{assertllm2}.
Column \textit{Proof} reports the percentage of generated assertions that are successfully proven for the given design, column \textit{COI} reports the \ac{coi} coverage, and column \textit{Formal} reports the formal coverage computed by JasperGold.
For the AssertLLM benchmark, our workflow performs best regarding proof rate, with a 2.3x improvement over AssertionForge for both the self-reported and re-run results.
As for formal coverage, our workflow performs best across both categories and baselines, where AssertionForge only self-reports \ac{coi} coverage.
On the AssertLLM2 benchmark, proof percentage is very consistent compared to AssertLLM for our workflow.
\ac{coi} coverage for our workflow is also very high, at over 90\%.
Formal coverage is where our proposed workflow achieves the best results compared to the baselines, performing 2.54x better than the re-run AssertLLM2 Plain.
In total, our evaluation demonstrates the capability of \acp{fvgherkinspecification} to serve as the foundation for \ac{fpv}, scoring very high across both \ac{coi} and formal coverage.

Across the two sets of experiments, our overall results demonstrate the suitability of our proposed end-to-end workflow to offer an integrated view on the use of \acp{llm} for \ac{eda}.
For both \ac{rtl} design and assertion generation, our workflow is able to outperform other established \ac{llm}-based methods in quality metrics such as functional correctness and coverage.
We acknowledge that some baselines were able to achieve results that are similar to those of our workflow, but we point out that the integration of enhancements like \ac{cot} prompting and \ac{rag}, or advanced techniques from these baselines into our workflow may lead to further improvements, e.g. utilizing additional agents for \ac{rtl} design like MAGE or constructing knowledge graphs for assertion generation like AssertionForge.
\section{Conclusion} \label{sec:conclusion}

In this work, we proposed an integrated view on the use of \acp{llm} for \ac{eda} through the establishment of an \ac{llm}-enabled behavior driven hardware development workflow.
We introduced \acp{fvgherkinscenario} for \ac{cnl} specifications as the foundation for the development of formally verified hardware designs.
By experimental evaluation, we confirmed that our proposed workflow can outperform established \ac{llm}-based methods for both \ac{rtl} design and \ac{fpv} based on \acp{fvgherkinspecification}.
Future work may investigate whether the integration of further options for the three agents leads to additional improvements of our workflow.

\section*{Acknowledgments}

This research has been supported by the German Ministry for Research, Technology and Space (BMFTR) with project ExaVerse (grant number 01IW25003).

\bibliographystyle{abbrv}
\bibliography{lit}

\end{document}